\documentclass[aps,prl,twocolumn,superscriptaddress,groupedaddress, floatfix]{revtex4}  
\usepackage{graphicx}  
\usepackage{subfigure}
\usepackage{mathtools} 
\begin{document}

\title{Rheology, diffusion, and velocity correlations in the bubble model}
\author{Arka Prabha Roy}
\author{Kamran Karimi}
\altaffiliation[Present address: ]{UJF Grenoble, France.}
\author{Craig E. Maloney}
\altaffiliation[Present address: ]{Northeastern University.  Boston, MA 02115}
\affiliation{Carnegie Mellon University, Pittsburgh, PA 15213}
\begin{abstract}
We present results on spatio-temporal correlations in the so-called mean drag version of the Durian bubble model in the limit of small, but finite, shearing rates, $\dot{\gamma}$.
We study the rheology, diffusion, and spatial correlations of the instantaneous velocity field.
The quasi-static (QS) effective diffusion co-efficient, $D_e$, shows an  anomalous system size dependence indicative of organization of plastic slip into lines along the directions of maximum shearing.
At higher rates, $D_e$ decays like $\dot{\gamma}^{-1/3}$.
The instantaneous velocity fields have a spatial structure which is consistent with a set of spatially uncorrelated Eshelby transformations. 
The correlations are cut off beyond a length, $\xi$.
$\xi\sim \dot{\gamma}^{-1/3}$ which explains the $D_e\sim\dot{\gamma}^{-1/3}$ behavior. 
The shear stress, $\sigma$, follows a similar rate dependence with $\delta\sigma=\sigma-\sigma_y\sim \dot{\gamma}^{1/3}$ where $\sigma_y$ is the yield stress observed in the QS regime.
These results indicate that the form for the viscous dissipation can have a profound impact on the rheology, diffusion and spatial correlations in sheared soft glassy systems. 

\end{abstract}

\maketitle

In recent years, it has become clear that shear flow in many types of amorphous solids -- structural glasses such as metallic or polymer glasses; soft glasses such as pastes, emulsions, or foams;  sheared granular matter -- is governed by localized plastic shear transformations~\cite{Falk:1998fx,Maloney:2006oc,Tanguy:2006nt,Jensen2014}.
These solids behave like elastic bodies below a yield stress, $\sigma_y$, but flow like viscous liquids above.
At low shearing rates, the response becomes bursty and intermittent, resembling other slowly driven out-of-equilibrium systems such as pinned elastic manifolds, Barkhausen noise in disordered ferromagnets, martensitic phase transformations, or dislocation-mediated crystal plasticity~\cite{Sethna:2001jh}.
One central question is how finite driving rate affects the spatio-temporal fluctuations and how this impacts the average response.

In the present case of amorphous solids, models have focused at the particle scale\cite{Maloney:2006oc,lerner:066109,Maloney:2008lj,Salerno2012,maloney:225502,Maloney:2004cu,Lemaitre:2007gl,lemaitre:065501,Chattoraj2011,Bailey:2007bl,Tanguy:2006nt,Tsamados:2009zr,Salerno2013} and at coarser, meso-scale levels~\cite{Bulatov1999a,Bulatov1999,Onuki:2003sh,Jagla:2007tb,Lin2014,ISI:000332617600014,Nicolas2014,Homer:2010oq,Homer:2009kl,Talamali:2012qf,Baret:2002fk,Rodney2011,Vandembroucq2011,Talamali:2011pd}.
At the meso-scale, one envisions plastic flow occurring due to localized yielding events that are observed in experiments and in the particle-scale models.
One may construct mean field theories based on these localized yielding events.
Some examples are the shear transformation zone (STZ) theory\cite{Langer2015}, the soft glassy rheology (SGR) model\cite{Sollich1996,Sollich1997}, and the Hebraud-Lecquex model (HL)\cite{Hebraud1998,Agoritsas2015}.
These models vary in their assumptions about the distribution of states, the barriers seen by the states, and the dynamics of the yielding events, but they all predict a so called Herschel-Bulkley (HB) power-law rate dependence of the shear stress where $\delta\sigma\sim\dot{\gamma}^\beta$.
However, the exponent, $\beta$, varies from model to model.

One can, alternatively, construct explicit real-space models. 
These so-called elasto-plastic models (EPMs) were pioneered by Bulatov and Argon and many variations have been proposed in the intervening years~\cite{Bulatov1999a,Bulatov1999,Onuki:2003sh,Jagla:2007tb,Lin2014,ISI:000332617600014,Nicolas2014,Homer:2010oq,Homer:2009kl,Talamali:2012qf,Baret:2002fk,Rodney2011,Vandembroucq2011,Talamali:2011pd}.
In most, with a few exceptions~\cite{Onuki:2003sh,Jagla:2007tb}, the loads are transferred instantaneously across all space as the site in question undergoes a yielding event.
In one of the few EPMs where loads \emph{are} transferred dynamically, a $\delta\sigma\sim \dot{\gamma}^{0.5}$ rheology was also observed~\cite{Jagla:2007tb}.
Recently, Liu {\it et. al.}~\cite{liu-condmat}, have shown that in a rate dependent EPM, the rheology exhibits a crossover from a non-trivial universal scaling regime at low rate where $\delta\sigma\sim \dot{\gamma}^{0.65}$ to a mean field behavior at higher rate where $\delta\sigma\sim \dot{\gamma}^{0.51}$.

Also at the particle scale, one generally observes HB rheology, often with a $1/2$ exponent. 
Molecular dynamics (MD) simulations on low temperature Lennard-Jones (LJ) glasses~\cite{lemaitre:065501} are in agreement with the HL model which predicts $\delta\sigma\sim \dot{\gamma}^{1/2}$.
In these simulations, there was also a power-law dependence on rate in the correlation length governing the plastic strain field and the diffusion coefficient, and a connection between these spatial correlations and the HB exponent was suggested. 
In foam simulations using Durian's bubble model with a linear drag imposed on relative velocity of particles~\cite{Langlois:2008ei} and more sophisticated models incorporating non-linear elasticity and non-linear drag between the particles~\cite{Seth:2011fu} along with corresponding experiments on similar systems~\cite{Nordstrom:2010fk,Seth:2011fu,Basu2014} all show behavior consistent with $\delta\sigma\sim \dot{\gamma}^{1/2}$. 

Between the particle-based simulations~\cite{lemaitre:065501,Langlois:2008ei,Seth:2011fu}, the EPMs~\cite{Jagla:2007tb, liu-condmat}, and the HL  model~\cite{Hebraud1998,Agoritsas2015}, one might expect $\delta\sigma\sim \dot{\gamma}^{1/2}$ to be universally applicable to many systems where the basic picture of local yielding and stress redistribution holds.
We show here that a simple variant of Durian's bubble model~\cite{Ono:2002gq,Ono:2003so} falls outside this class with an HB exponent of roughly $1/3$.
Furthermore, the class of experiments which are closest to this model -- amorphous Bragg-Nye bubble rafts -- ~\cite{Pratt2003,katgert:066318,Katgert2008,Mobius2008,Katgert2010} also show a HB rheology with an exponent consistent with $1/3$.

In this work, we study a version of the bubble model~\cite{Ono:2002gq,Ono:2003so} where the drag on the particles arises from motion with respect to the background suspending fluid; the so-called mean field drag (MFD) variant.
A mean-drag term could be appropriate for modeling systems such as particles at an air-water interface~\cite{Pratt2003,katgert:066318,Katgert2008,Mobius2008,Katgert2010} where dissipative forces on a particle may be governed by the generated subsurface flow. 
Early results on the MFD model were restricted to small systems and large shearing rates~\cite{Ono:2002gq,Ono:2003so}.
More recent studies have focused on the behavior near the jamming transition.
Here, we work with much larger systems and at much lower rates than previous studies and at volume fractions well above the random close packing point (precisely the same volume fractions and particle size and stoichiometry as studied by Ono {\it et. al.}~\cite{Ono:2002gq,Ono:2003so}).

We show that the MFD bubble model gives an HB exponent of roughly $1/3$.
Surprisingly, there are long range correlations present in the instantaneous \emph{velocity} fields that are of the classical Eshelby form one would expect for the \emph{displacement} fields induced by a local shear transformation in an elastic matrix~\cite{Picard:2004oe}.
The Eshelby correlations are cutoff beyond a length, $\xi$, which scales like $\xi\sim\dot{\gamma}^{-1/3}$.
$\xi$ governs the effective diffusion co-efficient, $D_e$, as in the MD simulations~\cite{lemaitre:065501}, with $D_e\sim\xi$, and the HB exponent via $\delta\sigma\sim 1/\xi$.
This gives an effective Stokes-Einstein relation: $D_e\delta\sigma=\text{const}$ as in the MD case~\cite{lemaitre:065501} (where we define $\delta\eta=\delta\sigma/\dot{\gamma}$).

We consider two dimensional (2D) system of soft disc particles in a bi-disperse mixture of equal number of small and large discs with diameter ratio, $D_S:D_L=1:1.4$~\cite{Ono:2002gq,Ono:2003so}.
We simulate this mixture at zero temperature using the bubble model introduced by Durian\cite{Durian:1995er,Durian:1997ph}. 
The harmonic interaction between discs $i$ and $j$ is, $U_{ij}=k_e\delta_{ij}^2/2$  if $\delta_{ij}<0$ and zero otherwise, where $\delta_{ij}=2r_{ij}/(D_i+D_j)-1$, is the overlap distance, $k_e$ is the elastic spring constant. 
Viscous dissipation is taken into account in a mean-field fashion with the total drag force on disc $i$, $\vec{F}_i^D=-b(\vec{v}_i-y_i\dot{\gamma}\hat{x})$, where $b$ is the damping parameter, $y_i$ is the location of the particle projected along the flow-gradient direction, $\hat{x}$ is the unit vector in the flow direction, and $\dot{\gamma}$ is the imposed shearing rate. 
Since the particles are considered massless, the dynamics are given by balancing elastic and drag forces.
The only timescale in the model is the timescale for elastic relaxation under the drag force: $\tau_D=\frac{b}{k}$; in our simulations we have used $b=1$ and $k=1$. 
We report all lengths in units of $D_S$ and all times in $\tau_D$. 
We define the volume fraction, $\phi$, as $\pi(N_L D_L^2 + N_S D_S^2)/4L^2$, where $N_L$ and $N_R$ are the number of big and small particles respectively, and $L$ is the simulation box size. 
We set $\phi=0.9$ which is far above the jamming point, $\phi_J\approx 0.843$ in two dimension. 
We report results for different sizes, $L= 40, 80, 160$ corresponding to a total number of, $N = 1240, 4960, 19840$ particles.
Lees-Edwards boundary conditions are used to impose simple shear flow~\cite{AllenTildesley}.
We refer to the flow direction as $x$ and the gradient direction as $y$. 
We use the standard Irving-Kirkwood expression for the shear stress $\sigma=(1/L^2) \sum_{ij} F_{xij}r_{yij}$~\cite{AllenTildesley}~\footnote{Note that we exclude the contribution from the viscous forces which is negligible in the regime of shearing rates studied here.}.
In this letter, we report on the steady state achieved beyond $50\%$ strain, for different $L$ and rates, $\dot{\gamma}$, and all statistics are taken between $50\%$ and $150\%$ strain.
All velocities discussed below are the non-affine velocities defined with respect to the background flow.
\begin{figure}[h]
\centering
	\includegraphics[scale=0.48]{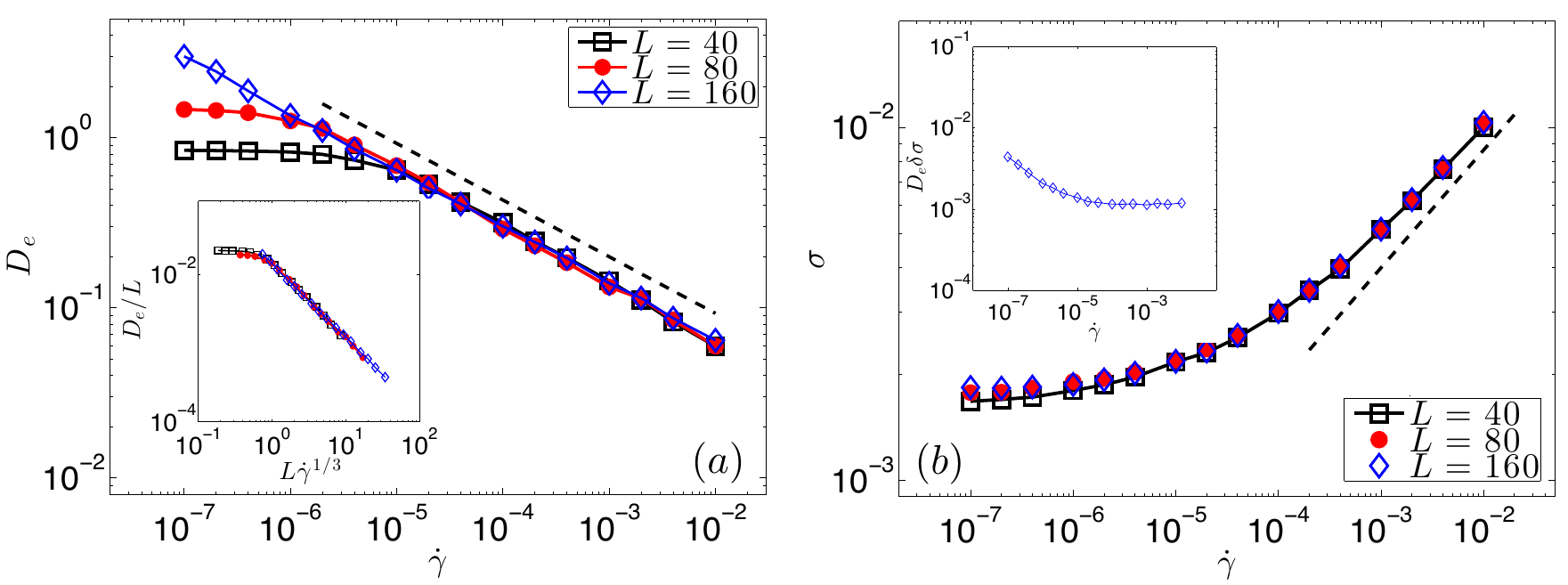}
	\caption{
	(a) $D_e$ vs. $\dot{\gamma}$ for systems of size $L=40$, $80$, and $160$; thin dashed line has a slope of $-1/3$. 
	Inset: $D_e/L$ vs. $L\dot{\gamma}^{1/3}$.
	(b) $\sigma$ vs $\dot{\gamma}$; thin dashed line has a slope of $1/3$. 
	Inset: effective athermal Stokes-Einstein temperature, $D_e\delta\sigma$, for the $L=160$ system for $\sigma_y=0.0011$.} 
	\label{fig:stress_diffusion}
\end{figure}

In Fig.~\ref{fig:stress_diffusion}a, we plot the effective diffusion constant $D_e\doteq \lim_{\Delta\gamma\rightarrow\infty} \langle \Delta y^2\rangle / 2\delta\gamma$ vs. $\dot{\gamma}$  for the three different system sizes, $L=40, 80$ and $160$.
$\Delta y$ is the particle displacement in the direction transverse to the flow for a time during which a strain of amplitude, $\Delta\gamma=\dot{\gamma}\Delta t$, was applied.
At large rates, $D_e$, is independent of $L$ and follows a $\dot{\gamma}^{-1/3}$ power law with remarkable precision and for over four decades of rate for the case of the $L=160$ system.
At the lowest rate in the QS regime, a plateau is clearly visible for $L=40$ and $80$.
In the inset, we plot $D_e/L$ against $L\dot{\gamma}^{1/3}$.
The data shows a good collapse, indicating that the QS diffusion, $D_{QS}\sim L$.
This behavior has been understood to arise from system spanning lines of slip in MD simulations of LJ glasses~\cite{lemaitre:065501,Lemaitre:2007gl,Maloney:2008lj,Chattoraj2011}.
The bubble model exhibits these same slip lines in the QS regime.
It is, rather, the rate dependence which is different here -- both in the rheology and diffusion.
Because of the relatively weak rate sensitivity, the $L=160$ system shows a strongly rate dependent $D_e$ even at $\dot{\gamma}=10^{-7}$, as it has not yet reached its QS plateau.

In Fig.~\ref{fig:stress_diffusion}b, we plot $\sigma$ vs. $\dot{\gamma}$.
At low rates, one approaches the yield stress, $\sigma_y$.
At higher rates, one approaches a power-law regime which is described reasonably well by $\delta\sigma\sim\dot{\gamma}^{1/3}$ for over a decade. 
We cannot rule out a cross-over to a different behavior at very low rates as one approaches the QS limit.
Regardless, the rate sensitivity is significantly less than $\dot{\gamma}^{1/2}$.
Ono {\it et. al.}~\cite{Ono:2003so} have already studied the diffusion explicitly and the rheology implicitly (via the mean-squared-velocity); but here we explicitly demonstrate the $\dot{\gamma}^{\pm 1/3}$ scaling and the finite size effects in the QS regime. 
In the inset, we plot $D_e\delta\sigma$ for the $L=160$ system.
$D_e\delta\sigma$ can be considered a kind of effective Stokes-Einstein temperature.
Our results for $D_e\sigma$ agree with Ono {\it et. al.} for the $\dot{\gamma}$ where the studies overlap, but we show here that one can obtain a remarkably constant value of $D_e\delta\sigma$ for over three decades in rate for the $L=160$ system.

\begin{figure}[h]
\centering
	\includegraphics[scale=0.48]{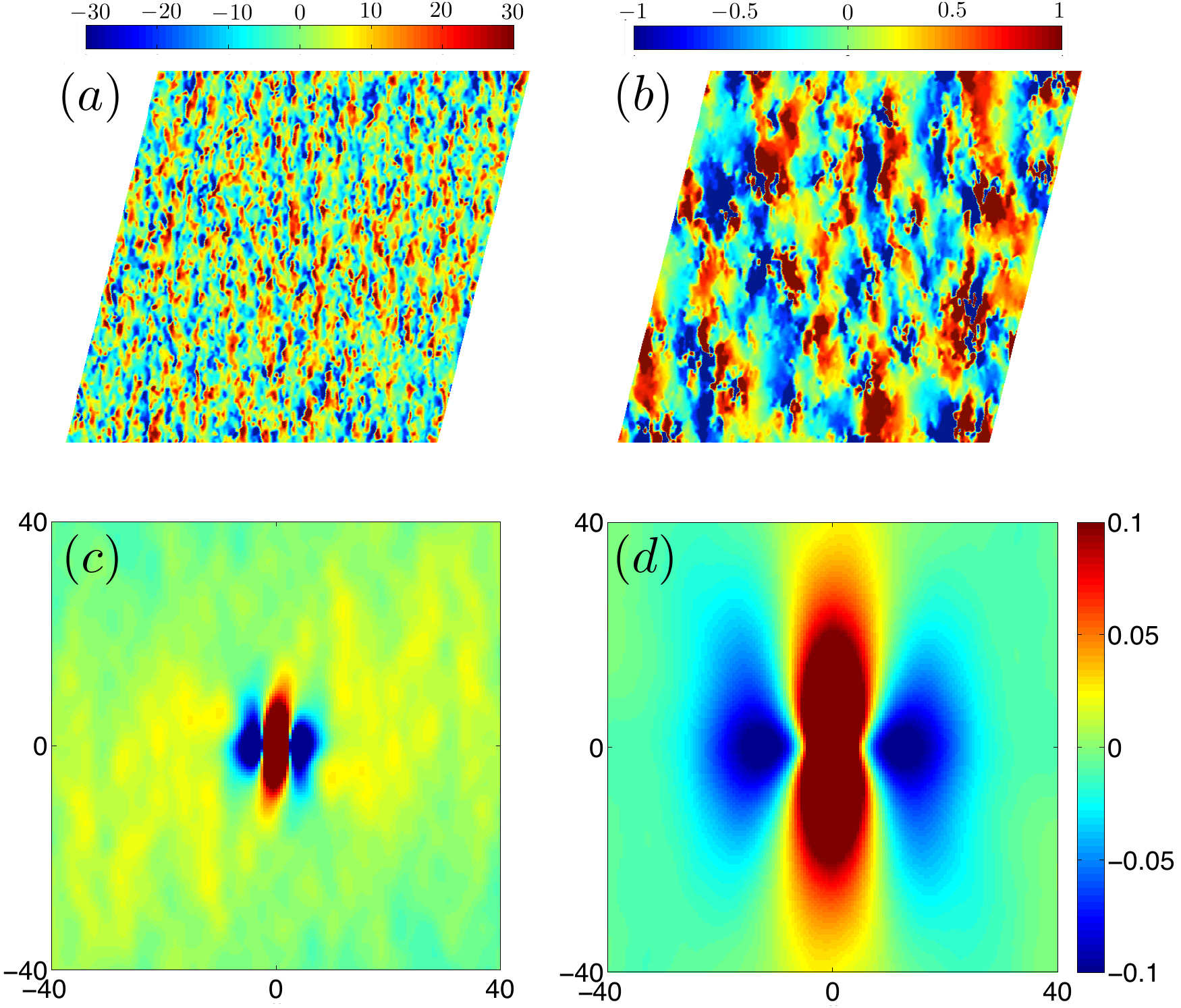}
	\caption{
	Top: A typical map of $10^{4}v_y$ for (a) $\dot{\gamma}=10^{-3}$ and (b) $\dot{\gamma}=10^{-5}$ using $L=160$.
	Bottom:  Autocorrelation of $y$-velocities $C_{v_y}(\vec{R})/C_{v_y}(x=1)$ for (c) $\dot{\gamma}=10^{-3}$ and (d) $\dot{\gamma}=10^{-5}$.} 
	\label{fig:yVelocity_realSpace}
\end{figure}
In Fig.~\ref{fig:yVelocity_realSpace}, we plot a typical snapshot of the $y$-component of the particle velocities at two typical rates, (a) $\dot{\gamma}=10^{-3}$ and (b) $\dot{\gamma}=10^{-5}$, and their respective time-averaged spatial autocorrelation functions, $C_{v_y}$, in (c) and (d).
As usual, we define $C_{v_y}(\vec{R})=\langle v_y (\vec{R}+\vec{r},t) v_y (\vec{r},t)\rangle_{(\vec{r},t)}$.
The large coherent spatial structures are obvious.
These correspond to long lines of particles which are all slipping together \emph{at the same instant}.
The sharp jumps from blue to red as one traverses the image from left to right indicate discontinuities in the velocity field with counter-clockwise vorticity.  
The real-space correlation functions reflect the visual impression.
The $y$-velocities have strong correlations along the $y$ direction.
They are correlated along the $x$-direction for some distance and eventually become anti-correlated.
One would naturally associate the cross-over from correlation to anti-correlation with the typical spacing between the slip lines.

\begin{figure}[h]
\centering
	\includegraphics[scale=0.48]{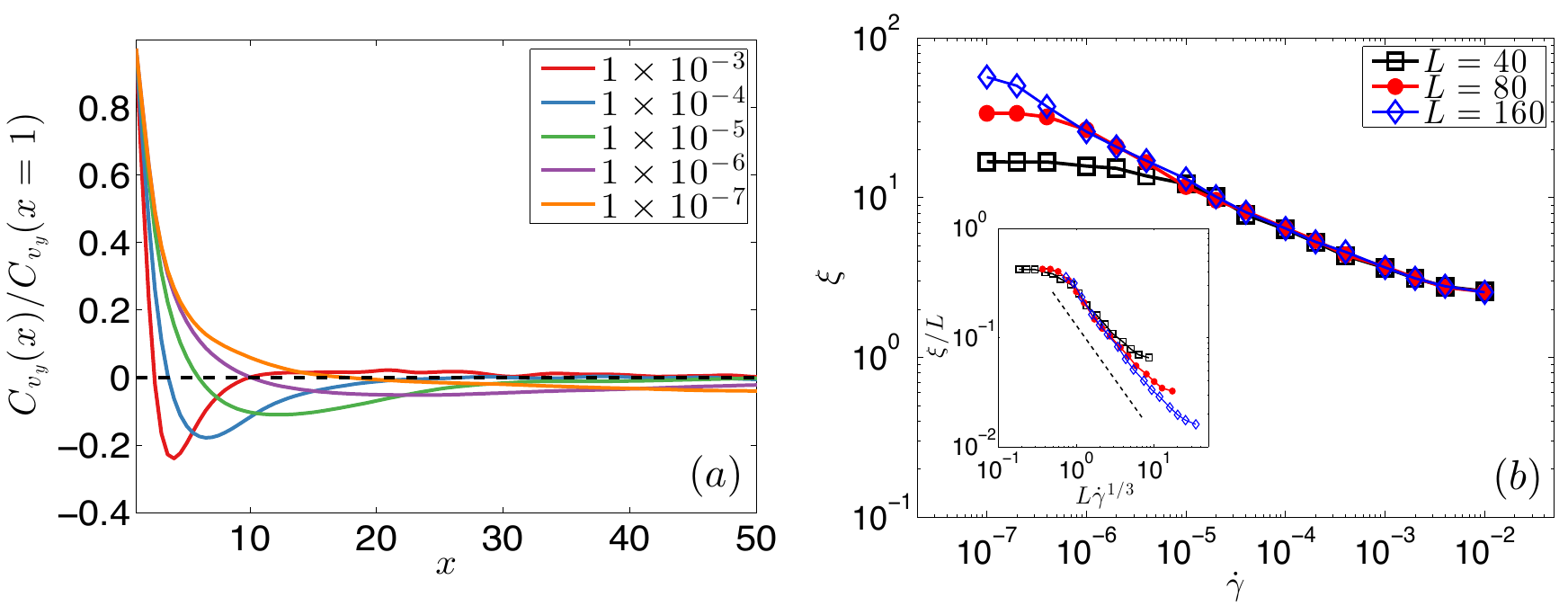}
	\caption{
	(a) $C_{v_y}(R=x)/C_{v_y}(x=1)$ for different $\dot{\gamma}$, $L=160$; black dashed line shows zero correlation. 
	(b) $\xi$ vs $\dot{\gamma}$ for different $L$. 
	Inset: $\xi/L$ vs $L\dot{\gamma}^{1/3}$; thin dashed line has a slope of $-1$.}
\label{fig:velCorTraceScaling}
\end{figure}
In Fig.~\ref{fig:velCorTraceScaling}a, we plot the traces of $C_{v_y}$, along the $x$-separations, normalized by the $x=1$ values, for various shearing rate for $L=160$.
In Fig.~\ref{fig:velCorTraceScaling}b, we plot the location, $\xi$,  of the minima of each $C_{v_y}$ curve as a function of rate for various system size.
As with the $D_e$ plots, for $L=40$ and $80$, we see a clear QS plateau at the lowest rates where $\xi$ saturates near the system size, while the $L=160$ system is just starting to show system-size dependent behavior at the lowest rate.
At higher rates, the data is well described by a $\xi\sim\dot{\gamma}^{-1/3}$ power law.
There are deviations from scaling in the high $\dot{\gamma}$, small $\xi$ regime, below about $\xi\approx 5$.
Nonetheless, we can observe over a decade of scaling for the $L=160$ system.
In the inset, we plot $\xi/L$ vs $L\dot{\gamma}^{1/3}$ showing the $\xi_{QS}\sim L$ quasi-static scaling~\footnote{The data cannot rule out a $\log{L}$ correction.}.

\begin{figure}[h]
\centering
	\includegraphics[scale=0.55]{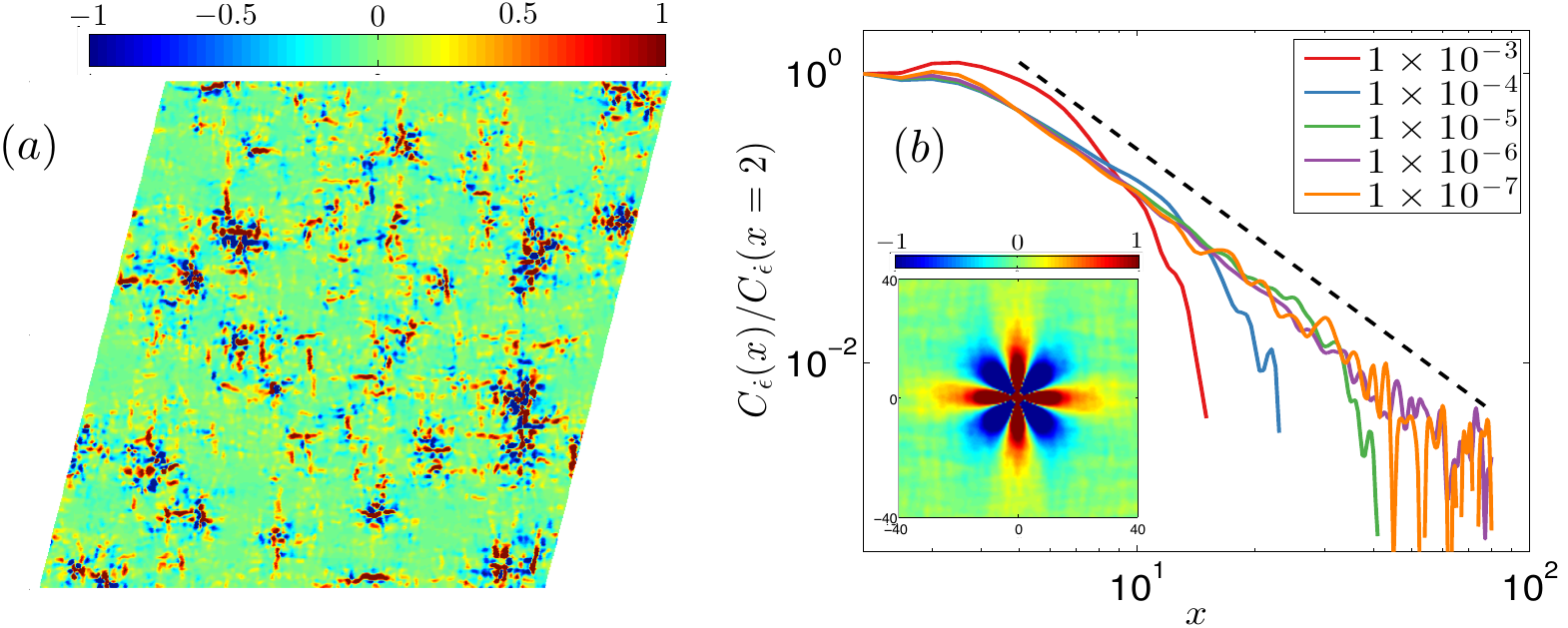}
	\caption{
	(a) A typical map of $10^4\dot{\epsilon}$ for $\dot{\gamma}=10^{-5},L=160$.
	(b) $C_{\dot{\epsilon}}(R=x)/C_{\dot{\epsilon}}(x=2)$ for different $\dot{\gamma}$, $L=160$; thin dashed line has a slope of -2.
	Inset: $10^2 C_{\dot{\epsilon}}(\vec{R})/C_{\dot{\epsilon}}(r=0)$ for $\dot{\gamma}=10^{-5}$.} 
	\label{fig:strainRate}
\end{figure}
In Fig.~\ref{fig:strainRate}a, we plot a typical snapshot of the symmetrized shear-strain-rate field, $\dot{\epsilon}\doteq(\partial_x v_y + \partial_y v_x)/2$, for $\dot{\gamma}=10^{-5}$ for precisely the same configuration as shown in Fig.~\ref{fig:yVelocity_realSpace}b.~\footnote{We define the strain rate by a finite difference of a linear interpolation of the particle velocities onto a fine mesh.} 
The vertically oriented slip lines with counter-clockwise vorticity, visible as sharp blue-red discontinuities in Fig.~\ref{fig:yVelocity_realSpace}b, appear now as red vertical lines.
Horizontally oriented slip lines with clockwise vorticity, which would have been impossible to detect in the $v_y$ fields, appear here as well. 
In real space, these slip lines appear to form a grid with a characteristic spacing consistent with $\xi$.

In Fig.~\ref{fig:strainRate}b, in the inset, we plot the respective time-averaged spatial autocorrelation function.
The angular variation has a quadrupolar symmetry reminiscent of the strain fields one would obtain from Eshelby inclusions~\cite{Picard:2004oe} with strong correlations along the directions of maximum shear and strong anitcorrelations $45$ degrees away.
We have checked that the angular dependence, at intermediate distances, is consistent with the $\cos{4\theta}$ dependence one would observe for an Eshelby transformation.

In Fig.~\ref{fig:strainRate}b, in the main plot, we plot the correlations along the directions of maximal shear, arbitrarily normalized to unity at $x=2$, for $L=160$.
For the highest rates with $\xi\leq 5$, corresponding to the regime with deviations from scaling observed in Fig.~\ref{fig:velCorTraceScaling}, the curves have a non-universal form.
At lower rates, where particle-size effects are less important, one observes a $C\sim r^{-2}$ power law for $x<\xi$.
The $r^{-2}$ power law is precisely what one would obtain for uncorrelated Eshelby transformations.

In summary, we have shown that the bubble model, with so-called mean field drag, exhibits surprising correlations in the instantaneous velocity fields.
These correlations are of precisely the Eshelby form in the limit of low shearing rate.
Strong correlations -- if not precisely of the Eshelby form -- were to be expected at low shear rate in the finite time displacements, in analogy with MD simulations, but it was surprising to see them here in the velocity field and surprising to see them in precisely the Eshelby form.
Although we find essentially the same connection between the correlation length and effective diffusion co-efficient as the MD simulations of Lemaitre and Caroli~\cite{lemaitre:065501}, both the rate dependence of the correlation length and the connection between that length and the rheology is qualitatively different here.

In~\cite{lemaitre:065501}, a length, $l$, was introduced to rationalize the rate dependence of $D_e$.
The relation $D_e\sim \xi$ (for $\xi << L$), remains the same here and can be understood as a simple kinematic consequence of plastic deformation essentially organizing along lines as in ref~\cite{lemaitre:065501}.
However, the rate dependence of $\xi$ is different: here it scales like $\dot{\gamma}^{-1/3}$ whereas, in reference~\cite{lemaitre:065501}, $l\sim\dot{\gamma}^{-1/2}$.
The argument in~\cite{lemaitre:065501} for $\xi\sim\dot{\gamma}^{-1/2}$ was quite general and relied essentially only on: i) the assumption of a $\dot{\gamma}$ independent timescale, $\tau$, for elementary shear transformations and ii) a basic picture of elementary shear transformations releasing a characteristic, $\dot{\gamma}$ independent, strain (or, equivalently, stress).
It will be important in the future to try to understand precisely why this argument breaks down for the MFD bubble model.

Furthermore, the connection between $\xi$ (or, equivalently $D_e$) and $\delta \sigma$ is different than in~\cite{lemaitre:065501}.
Lemaitre and Caroli argued that
a characteristic time for stress relaxation, $\tau$, scaling linearly with the correlation length along with the observed $\xi\sim\dot{\gamma}^{-1/2}$ behavior could explain the $\delta\sigma\sim\dot{\gamma}^{1/2}$ rheology.
Here, since we observe $\xi\sim\dot{\gamma}^{-1/3}$ and $\delta\sigma\sim\dot{\gamma}^{1/3}$, we would need to invoke a $\tau\sim\xi^2$ rather than $\tau\sim\xi^1$ relation to explain the connection
.

Interestingly, we observe a rate independent effective athermal Stokes-Einstein relation connecting the diffusion to the rheology: $D_e\delta\sigma=D\delta\eta=\text{const}$.
Note that our results agree with Ono {\it et. al.}~\cite{Ono:2003so} for $D_e\sigma$, but we find it interesting that $D_e\delta\sigma$ remains constant over a much broader range of shearing rates than $D_e\sigma$.
Of course, at larger rates, $\dot{\gamma}\gtrsim 10^{-2}$, where $\xi\approx\mathcal{O}(1)$ one would expect an increase in $D_e\sigma$, as observed in~\cite{Ono:2003so}, but we are not interested in this regime for purposes of this study.

It seems to us that the emergence of a rate independent effective Stokes-Einstein temperature is a more compelling -- and apparently general -- connection between the diffusion and rheology than a particular relationship between correlation length and stress relaxation time as in~\cite{lemaitre:065501}.
We further speculate that $D_e\delta\sigma$ may be related to various effective temperatures which arise in the STZ, SGR, and HL models.
In particular, Langer argues~\cite{Langer2015},  on very general grounds, that the effective temperature should become rate independent at low rate whenever the timescale of the external drive is much longer than any internal relaxation times.
This rate independent $T_{\text{eff}}$ would then give the connection between the diffusion and rheology via an effective athermal Stokes-Einstein relation~\cite{Ono:2002gq,Ono:2003so}. 


In addition to the questions about the origin of the rate dependence of $\xi$ and the relation between $\xi$ and $\delta \sigma$, there are some other obvious open questions.
i) If one imposes a drag on local shearing rates -- as in the original version of the bubble model -- instead of velocities, does one observe the same rate dependence on $\dot{\gamma}$ of $\delta\sigma$, $D_e$, and $\xi$?
We are in the process of conducting these studies now.
ii) Does one recover the same scaling laws when physical stress propagation -- in terms of the magnitude and type of drag -- is implemented in the EPMs?
We note that Jagla's model~\cite{Jagla:2007tb}, which includes physical stress propagation, in its current form, damps the local strain rate, as in the original formulation of Durian's bubble model, rather than the non-affine velocity, as in the present MFD version.
It would be quite interesting and relatively straightforward to modify this model to implement velocity damping rather than strain-rate damping to see if one would obtain the $\dot{\gamma}^{\pm 1/3}$ scaling laws.
iii) Do the $\dot{\gamma}^{\pm 1/3}$ scaling laws hold in 3D?
These are some of the questions which will need to be addressed on the road to understanding the rate dependent response of various athermal, sheared amorphous materials.

\acknowledgements
This material is based upon work supported by the National Science Foundation under Award Numbers DMR-1056564, CMMI-1250199, and PHY11-25915.
We acknowledge useful discussions with participants during the workshop on ÒAvalanches, Intermittency, and Nonlinear Response in Far-from-Equilibrium SolidsÓ at the Kavli Institute for Theoretical Physics, University of California, Santa Barbara and thank KITP for support during the course of this work.

\bibliography{master-mendeley}
\bibliographystyle{prsty}
\end{document}